\documentclass{amsart}
\usepackage{graphicx}
\begin{document}

\title {Damage Thermodynamics of Quasibrittle Materials}

\author{Leonid S. Metlov}

\address{Donetsk Institute for Physics and Engineering of NASU, 83114,.
Ukraine, Donetsk, str. R. Luxembourg, 72}

\email{metlov@mail.donbass.com}
\email{metlov@atlasua.net}

\thanks{This work was supported by Donetsk Innovation Center
(provision of computer and internet).}

\thanks{I express my gratitude to
Dr. V. Yurchenko the discussions with who permit me to  more
correct present my findings.}



\date{}

\dedicatory{}

\commby{}

%%% ----------------------------------------------------------------------

\begin{abstract}
The description of the early stage of microfracture growth in a
quasibrittle solid with thermodynamic positions is considered.
From the most general thermodynamic performances the "principle"
of minimization of free energy is received. The account high (down
to thirds) degrees in expansion of free energy concerning
parameters of a mechanical field and average microcrack energy has
allowed to write down the state equation for solids with
microcracks, to find equilibrium and nonequilibrium values of
their average energy.  From positions of a developed formalism the
reason of high stability of work developments on small depths and
mechanism of loss of their stability on the large depths is
explained. From the fact of presence of qualitatively various
behaviour of a material on the large depths the additional
estimated connection between parameters of the theory is
established.
\end{abstract}

%%% ----------------------------------------------------------------------
\maketitle
%%% ----------------------------------------------------------------------

\section*{Introduction}

Fracture phenomena are most complicated and intrigued fundamental
problem of the present-day. As of now, a sequence of fresh
approaches to attacking of this problem is outlined. Statistical
mechanics of cracks is most rigorous and sequential from them
\cite{BS96,Set01}, but very hard at a times. It is used out and
developed the methods of the complex systems, methods of the
percolation and fractal theories and setera \cite{Mish97}. Next
way for consideration of the problem is using of purely mechanic
descriptions \cite{Baza94,Perl} and non-equilibrium thermodynamics
\cite{Brady,Wata93}. Very interest and promising results are
obtained with the phase-field model
\cite{AKV00,ESRC01,KMMKK00,Sh96,ZBLH97} based on the phase
transition concept \cite{LL80}. In the phase-field models a order
parameter describes microscopic distribution of mass density in he
transition zone between a solid (phase I) and a separate crack
(phase II).

It is known approaches for describing of the fracture problem as
the first-order phase transition \cite{ZRSV96,VZS96}. One may
regard an each separate crack as nucleus or droplet of new phase
with determinate activation energy. Such approaches are more
suitable for description of ideal zero-defects solid, for which
the activation has thermal character and activation energy haven
enough high value is fixed.

In the other side it is known opposite approaches for the same
seemingly problem with position of the second-order phase
transition \cite{MGP01}. Real structural inhomogeneous solids have
the multitude manifest and latent defects in its bulk. They may be
activated at different energies from zero for manifest defects to
theoretical limit for ideal solid. Thus one may suppose that
quantity of defects in a solid rests constant at all times, and to
describe degree its activation by average energy per one defect
\cite{MM01,Wata93}. Such "order parameter" describes mesoscopic
level state, a elementary unit of which consist multitude
structural inhomogencities.

Practically relaxation of quasibrittle solid with the plastic
(viscous) mechanism even at slow deformation is too hard owing to
large rigidity of their molecular organisation. Therefore,
probably, the unique mechanism of relaxation in this case is
activation and growth of microcracks, their further merge and
branching (damage) and, at the end, failure thereof. The
destruction of quasibrittle solid with this mechanism enough
frequently is naturally observed, for example at deformation of
rock around of a underground working on too large depth
\cite{Bakl88,Rup75,StPr92}.

In present paper the behaviour of quasibrittle solids at high
loading (deformation) is considered from general thermodynamic
positions. The applicability of performances of equilibrium
thermodynamics in this case is limited to intermediate temporary
scale smaller necessary for diffusion healing of microcracks. From
consideration are excluded both infinite timeout, and may be
processes of microcrack origin. It is considered, that a real
fragile solid already has natural loosenings, defects etc., which
are only made active by an applied external mechanical field.

It is known the destruction is irreversible thermodynamic process.
In classical thermodynamics irreversibility and
non-equilibribility are taken as interrelation notes \cite{Bazar}.
That is any irreversible process is non-eqilibrium, and conversely
any non-equilibrium one is irreversible. It is true for infinite
slow processes. In real situation of intermediate time there is
possibility when process has irreversible character, but one is
equilibrated or more precisely quasiequilibrated.

\section{Constitutive relations}

The occurrence of microcracks (damage) conducts to growth of
internal energy of a solid on value of work of external force
equal to total energy of the broken off bandings. In this case for
slow equilibrium processes basic equation of thermodynamics is
possible to write down as:

\begin{equation}\label{a1}
    du=Tds+\sigma_{ij}d\varepsilon_{ij}+ \varphi d \nu ,
\end{equation}

where $u$ - is density of internal energy, $J/m^{3}$, $T$ - is
temperature; $s$ - is density of entropy, $J/degree*m^{3}$,
$\sigma_{ij}$ - is tensor of stress, $N/m^{2}$; $\varepsilon_{ij}$
- is reversible (elastic) part of deformation tensor; $\varphi$,
$\nu$ - is accordingly average additional energy of system per one
microcrack, $J$, and thermodynamically paired quantity, connected
to it having dimension of microcrack density, $m^{-3}$. The
relationship (1) can be presented in other, more known form

\begin{equation}\label{a2}
    du=Tds+\sigma_{ij}d\varepsilon_{ij}+
    \sigma_{ij}d\varepsilon^{n}_{ij},
\end{equation}

where $\varepsilon^{n}_{ij}$ - tensor of inelastic deformations.
From the given expression it is visible, that "pumping" of
internal energy is carried out through an irreversible part of
work of a external stress. As other channels of dissipation are
not considered, this part of work can go on "pumping" of internal
energy in the form or thermal movement (that is taken into account
by first term in (1)), or in the form surface energy and energy of
microcracks interaction (that is taken into account by last term
in (1)). The form of the relationship (1) is more convenient, if
the research of a model of a solid are limited to energetic level
of consideration, that, however, it is quite enough for reception
of many useful and common results. If necessary of more detailed
analysis between macroscopic parameters $\sigma_{ij}$,
$\varepsilon^{n}_{ij}$ and internal ones $\varphi, h$ is necessary
to establish additional connection similar those in the
thermodynamic theory of dislocations \cite{KR56}.

In essence, the relation (1) is tantamount to stating that
irreversible part of external work is substituted with effective
external microcrack flow in a manner like thermal flow. Motivation
for such presentation is concluded that information about solid
state in form of inelastic deformation is lost for system. In
other side this information residues in the system in the form
it's changed microcrack structure. In this context there is quite
deep analogy between first and last terms of expression (1).
Namely, temperature $T$ has meaning of average energy per one
particle rather per one degree of freedom, while entropy has
meaning of measure of such effective degree of freedom. Parameter
$\varphi$, as reported above, has meaning of average additional
energy per one defect (in our case, microcrack), that is it plays
a role of static "temperature". Logically to surmise that density
of microcracks is some analogue of static "entropy". Really, when
system transits from a nonequilibrium state to it's equilibrium
one both entropy and density of microcracks have similar feature:
namely at suit external conditions each of them tends to
increasing. It is necessary to note, that the additional internal
energy connected to occurrence of a separate microcrack, is
proportional to the sizes by last, and the increase of average
energy of microcracks is equivalent to increase of their average
size. This additional energy is caused by direct break of banding,
formation of a free surface limiting a microcrack, and
reorganization of nuclear layers near the surface with arising of
additional internal stress.

Free energy of a system is such part of it's internal energy,
which is able to perform mechanical work. Unavailable energy is
that lost part of internal energy, which isn't able to perform
mechanical work. The thermal part of unavailable energy is well
known and equal to $T s$. It is clear, that part energy
accumulated in the microcracks is lost for work too. In the
context of said above let this part of unavailable energy be equal
to $\varphi \nu$. Then new expression for density of free energy
first has been wrote in the form:

\begin{equation}\label{a3}
    f=u-Ts-\varphi \nu ,
\end{equation}

so, that its complete differential can be written down in the form

\begin{equation}\label{a4}
    df=-sdT+\sigma_{ij}d\varepsilon_{ij}- \nu d\varphi ,
\end{equation}

    Ignoring by thermal contribution one may rewrite down (4) as

\begin{equation}\label{a5}
    df=\sigma_{ij}d\varepsilon_{ij}- \nu d\varphi ,
\end{equation}

From a relationship (5) follows, that the free energy is
potential, if variables $\varepsilon_{ij}, \varphi$ are chosen as
arguments, i.e. $f=f(\varepsilon_{ij}, \varphi)$. If the explicit
arguments dependence of free energy density was known, variables
$\sigma_{ij}$ and $\nu$ thermodynamically connected with arguments
could be determined by simple differentiation:

\begin{equation}\label{a6}
\sigma_{ij}=\frac{\partial f}{\partial \varepsilon_{ij}},
\nu=-\frac{\partial f}{\partial \varphi}.
\end{equation}

It is necessary to note, that on sense of the relationship (1),
the relationships (6) should express external influences on
elementary unit. At the same time parameter $\nu$ can make sense of
external parameter only for the mobile defects capable to move
through external interface of elementary unit. If the defects are
motionless, and can only arise or disappear inside volume, that,
strictly speaking, it is necessary to put $\nu=const$. Without the
loss of generality it is believed that $const=0$. Validity of some
extreme "principle" from here follows \cite{ILF89,LL80}:

\begin{equation}\label{a7}
    \frac{\partial f}{\partial \varphi}=0.
\end{equation}

By this meant that in the equilibrium stationary state the
irreversible work isn't carried out. If one exerts additional
constant stress new microcracks are arisen (activated) in the bulk
for as long as new equilibrium state is achieved. In the other
side condition (7) is meant the free energy has minimum in any
equilibrium state as $\varphi$ is conceptually internal parameter.

\section{Expanding at power series}

As the definition of an explicit expression for free energy
density is a hard task, we apply standard in such cases reception,
namely, we shall expand free energy on its arguments:

\begin{equation}\label{a8}
    f(\varepsilon_{ij}, \varphi)= f_{0}-\nu_{0} \varphi + \frac{1}{2} a
    \varphi^{2}-\frac{1}{3}b \varphi^{3}+ \frac{1}{2} \lambda
    \varepsilon^{2}_{ii}+\mu \varepsilon^{2}_{ij}-g \varphi
    \varepsilon_{ii} + \frac{1}{2}\overline{\lambda}\varphi
    \varepsilon^{2}_{ii} + \overline{\mu} \varphi
    \varepsilon^{2}_{ij} + e \varphi^{2} \varepsilon_{ii}+...
\end{equation}

First term $f_{0}$ is free energy of the non-loaded body in
absence of microcracks. The following three terms describe
features of dependence of free energy from microcracs in absence
external mechanical load. Term $\nu_{0} \varphi$ is meaningful to
bulk density of superficial energy of the completely isolated and
not interacted among themselves microcracks ( $\nu_{0}$ - is their
natural density). Square term with $\varphi$ describes dimensional
effect taking into account that fact, that the growth of energy of
a microcrack with increase of its size is slowed down. Cubic term
with $\varphi$ takes into account an opportunity of merge of
microcracks, when the average energy per one merged microcrack,
grows at common reduction of energy of system (sign "minus" at
$b>0$).

Terms $\lambda \varepsilon^{2}_{ii} / 2$  and $\mu
\varepsilon^{2}_{ij}$  in (8) describe decomposition of an elastic
part of energy through invariant of deformation tensor in a
continuous solid (in absence of microcracks).

Subsequent mixed terms have double interpretation: from the point
of view of influence of microcracks on deformation field and, on
the contrary, from the point of view of influence of deformation
field on behaviour of microcracks. Term $-g \varphi
    \varepsilon_{ii} $, on the one hand, takes
into account loosening up (increase of volume) non-loaded material
at presence of microcracks, and, on the other hand, change of
"own" energy of microcracks owing to work of mechanical forces.

Cubic mixed terms $\overline{\lambda}\varphi
    \varepsilon^{2}_{ii} / 2$ and $\overline{\mu} \varphi
    \varepsilon^{2}_{ij}$ also describe increase or reduction of
free energy at the expense of the uncoordinated deformation of
various elementary volumes, for which size of average energy of
microcracks $\varphi$ (or their average size) is a measure of such
incoordination. As a matter of fact, these terms describe
additional interaction of elementary volumes, however only
formally they can be treated as effective interaction of
microcracks. The given interaction is carried out through
far-acted field of macroscopic deformations. It is need to
distinguish such interaction from interactions of own deformation
field of microcracs, which decreases quickly enough \cite{Bar61}
and can play an appreciable role only at large density of
microcracks. The parameters $\overline{\lambda}, \overline{\mu}$
in this context take into account change of stiffness of a
material for the account of microcracs. Discussed terms are
possible to interpret and as influence on energy of a microcrack
of deformation field owing to the same uncoordinated deformation.

At last, term $e \varphi^{2} \varepsilon_{ii}$ describes, on the
one hand, smaller growth of a degree of material loosening with
growth of the microcrack sizes, with another - influence of
interaction through own deformed field. Notice that the centre of
the expansion may be not only zero as in (8), but any other value
of $\varphi$.

Leaving in (8) explicit dependence only from average energy of
microcracks, we receive:

\begin{equation}\label{a9}
    f(\varphi)= \overline{f_{0}} - \overline{\nu} \varphi + \frac{1}{2}
    \overline{a} \varphi^{2} - \frac{1}{3} b \varphi^{3} +
    ...,
\end{equation}

where

\begin{eqnarray}\label{a10}
  \nonumber
  \overline{f_{0}}=f_{0}+\frac{1}{2} \lambda
    \varepsilon^{2}_{ii}+\mu \varepsilon^{2}_{ij}  \\
   \overline{\nu}= \nu_{0} +g \varepsilon_{ii} -\frac{1}{2}\overline{\lambda}
    \varepsilon^{2}_{ii} - \overline{\mu}  \varepsilon^{2}_{ij}   \\
 \nonumber \overline{a}=a + 2 e \varepsilon_{ii} .
\end{eqnarray}

The diagram of free energy generally represents a cubic parabola,
the variants of possible forms of which and arrangement of them
concerning of coordinate axes at various values of tensor
components of deformation are given in the fig. 1. Three forms of
a cubic parabola - with two extremes on the diagram, with a point
inflexion and monotonously falling down diagram (fig. 1a) are
generally possible. From extreme "principle" (7) the equation for
definition of equilibrium values of average energy follows

\begin{equation}\label{a11}
    f'(\varphi)=- \overline{\nu} + \overline{a} \varphi - b
    \varphi^{2} = 0 .
\end{equation}

The decisions of the equation

\begin{equation}\label{a12}
    \varphi_{1/2}=\frac{1}{2b}(\overline{a} \pm
    (\overline{a}^{2} - 4 b \overline{\nu})^{1/2})
\end{equation}

are those, that is valid $\varphi_{1} > \varphi_{2}$. The
stability of roots of the equation (11) is determined by sign of
the second derivative of free energy:

\begin{equation}\label{a13}
    f''(\varphi_{1/2})=\overline{a} - 2 b \varphi_{1/2} = \mp
    (\overline{a}^{2} - 4 b \overline{\nu})^{1/2}) ,
\end{equation}

The first root is unstable, second one is steady and determines
some equilibrium value of average energy come on one microcrack.
As the state of a solid with microcracks, strictly speaking, is
nonequilibrium, but can exist long enough, it as a matter of fact
is metastable. The extreme principle in the form (11) determines
conditions of existence of its metastable state. If the deviation
of the current value of average energy from equilibrium is
insignificant, the solid keeps the integrity . Here it is meant,
that the current value of average energy of microcracks can, both
lag behind, and outstrip equilibrium value  , which with (12)
follows change of a mode of elastic loading of a sample. If the
current value of average current average energy is those, that  ,
the system passes on a falling down branch of free energy (curve
1, fig. 1a), that is equivalent to creep or later to macroscopic
destruction of a solid or, at least, to transition of destruction
to the following structural level (merge of microcracks).
Existence of a potential barrier separating a steady state from of
the falling down branch, depends from sign of the discriminate:

\begin{equation}\label{a14}
    D=\overline{a}^{2} - 4 b \overline{\nu} .
\end{equation}

At $D=0$ (curve 2, fig. 1a) the potential barrier is absent. In
this case body begins to creep at slightest change of parameter
$\varphi$. At negative values of discriminant (curve 3, fig. 1a)
the equilibrium values are absent at all, and the solid creeps
constantly. Under condition $D>0$ it takes place two different
decisions, between which should be observed an inequality
$\varphi_{1} > \varphi_{2}$. Increasing deformation of a material
so that to pass from negative values of discriminant $D$ to
positive them, it is possible to transfer a solid from a
conditionally equilibrium state to nonequilibrium state. In result
the value of average energy of microcracks will grow continuously
(creep), that will be achieved for the account, both growth of the
sizes of separate microcracks, and their merge at general downturn
of energy of system. Value $D=0$ separates regions of equilibrium
states and creep. Further, varying the free energy on deformation,
it is possible to determine a field of pressure:

\begin{equation}\label{a15}
    \sigma_{ij}=\frac{\partial f}{\partial \varepsilon_{ij}} =
    [(\lambda_{ef} \varepsilon_{ii}-g \varphi)\delta_{ij} -
    2 \mu_{ef}\varepsilon_{ij} ] ,
\end{equation}

Where

\begin{eqnarray}\label{a16}
  \nonumber
  \lambda_{ef}=\lambda + \overline{\lambda} , \\
   \mu_{ef}=\mu + \overline{\mu}
\end{eqnarray}

As it is visible, the elastic modules is varied of depending
on presence of microcracks. The positive values of parameters
$\overline{\lambda}$ and $\overline{\mu}$  ($\overline{\lambda}>0$
and $\overline{\mu}>0$) correspond to growth of effective stiffness
of a material with growth of average microcrack energy. The last
takes place, if during deformation, despite of break of a part
of the bindings the rest of them are capable to keep additional
action promoting more hard response. Growing of average microcrack
energy may be accompanied by strengthening of rest bonds with
increasing of internal stress around a microcrack. If it not so,
the break of bindings will cause downturn of rigidity of a material,
and the inequalities $\overline{\lambda}<0$ and $\overline{\mu}<0$
will be carried out. The first case is quite exotic, but the letter
one is thermodynamically more correct.

\section{Transition through zero}

As to metastable states correspondingly dependence of sign of the
parameter $\varphi_{2}$ three qualitatively various situations are
possible. The positive values of this parameter (curve 3, fig. 1a)
describe a real physical state of a solid. The negative values
(curve 1, fig. 1a) have no physical sense, and the free energy
formally does not depend in any way on microcracks. The
independence of free energy of parameter $\varphi$  in the field
of its negative values is represented by horizontal lines (fig.
1a). The equality to zero of this parameter determines the instant
of disappearance (in statistical sense) microcracks for the reason
that their sizes at zero their average energy are equal to zero
(collapse).

We shall consider features of transition of system through zero
of equilibrium parameter $\varphi_{2}$. Near to this point the
large powers in expansion of free energy it is possible to reject
and to be limited only to square-law approximation. Then the point
$\varphi_{2}=0$ is achieved at deformation with hydrostatic compression:

\begin{equation}\label{a17}
    \varepsilon^{(0)}_{ii} = -\frac{\nu}{g}<0 ,
\end{equation}

To the left of a point $\varphi=0$ the free energy, as was told above,
does not depend from variable $\varphi$  and, hence, it's derivative
of all orders on this variable identically address in zero. To the
right of a point $\varphi=0$ derivative from free energy are not
equal to zero and can be determined by double differentiation of
expression (8). Thus, second derivative of free energy will have
discontinuity at transition of a point $\varphi=0$:

\begin{eqnarray}\label{a18}
\nonumber
  f''_{\varphi} = \overline{a} = 0 ....if.... \varepsilon_{ii} >
    -\frac{\nu}{g}<0 , \\
  f''_{\varphi} = 0 ....if.... \varepsilon_{ii} < -\frac{\nu}{g}<0 ,
\end{eqnarray}

And also

\begin{eqnarray}\label{a19}
\nonumber
    f''_{\varphi,\varepsilon_{ii}} = -g > 0 ....if....
    \varepsilon_{ii} > -\frac{\nu}{g}<0 , \\
    f''_{\varphi,\varepsilon_{ii}} = 0 ....if.... \varepsilon_{ii}
    < -\frac{\nu}{g}<0 ,
\end{eqnarray}

In second by derivative on deformation the discontinuity is
absent, but the break on the diagram of dependence from parameter
$\varphi$ takes place. The behaviour and character of two
derivatives free energy discontinuity testifies that we have
to deal with some analogue of a second kind phase transition.

\section{"Elastic-creep" transition}

The presented here theory can be applied to qualitative
explanation of a phenomenon of destruction of rock around of
underground workings. It is known, that ones carried out on small
depths have enough large stability. Around workings carried out on
large depths the phenomena of the abnormal large displacement of
rock are observed. This is cause fast reduction of section of
working and, eventually, its complete destruction \cite{ZZCh73}.
According to the theory at small deformation, $\varepsilon_{ii}$
this is at small depths (the estimates are given below) for rock
around working, will be satisfied condition of existence of an
equilibrium state (case $D<0$), which is described by free energy
with one minimum. (curve 1 in the fig. 1a). At increase of shear
deformations (the bulk deformation in elastic approximation is
equal to zero), this is at the large depths, the character of
dependence of free energy will change, and it will be described
already by curve which is not having extremes (a curve 3 on the
fig. 1a). In this case system has no equilibrium states (at $D>0$)
and, hence, will be constant to creep.

Own to creep system evaluates in the side of growing of average
microcrack energy. After a time system lands in region of a large
 $\varphi$ and power expansion of free energy (8) becomes not true. The
state becomes exclusively non-equilibrium and new assumptions are
need for description of system. For example, one can assume that
creep changes medium parameters with rate in proportion to
deviator of stress tensor \cite{MM97} and so on.

The described above phenomenon prompts a graceful way of an
establishment of estimated connection between parameters of the
theory. Let's assume, that depth $H$ of a working, since which the
specified phenomenon begins to be shown, is equal 200m. The
hydrostatic pressure, appropriate to it, makes $\gamma H$, and
tangential tensor component of pressure on a contour of the
working is $2 \gamma H$ (where $\gamma = \rho g$, $\rho$ - is
density of rock equal on the average $2.6*10^{3} kg/m^{3}$, $g$ -
is free fall acceleration). From here $\sigma_{\tau}=10.4 MPa$.
Believing, that shear stiffness of a rock $30 GPa$, for shear
components of deformation tensor we receive value of the order
$\varepsilon_{\tau}=\sigma_{\tau}/\mu=0.00035$. The square of this
value has the order second tensor invariant of deformations
$\varepsilon^{2}_{ij}$ appearing in expanding of free energy (8).
Taking into account, that on contour of the working first tensor
invariant of deformation is identically equal zero (it follows
from exact decision of the problem in elastic production), from
condition of equality zero of a determinant (14) is possible to
write down as:

\begin{equation}\label{a20}
    a^{2} - 4 b (\nu_{0}-0.00035 \overline{\mu})=0.
\end{equation}

The given connection between parameters of the theory, alongside
with attraction of results of special experiments, can be useful
to numerical estimate. At the present moment the complete complex
of such experiments directed on measurement of parameters $a, b,
\nu_{0}, \overline{\mu}$,  and other parameters of the theory, to my
mind, are absent. The offered here consideration can serve as
stimulus for development of experimental development in this
direction.

\section{Practical sequels}

From the above an original idea of increasing of opening stable
in the Mining field is followed. The much troubles with traditional
ways for opening timbering is - one combats with no cause, but
with effects.

Usual evolution of natural fracture of rock around opening passes
through such stages as a) accumulation of microcracks and
b) macroscopical fracture

The first stage has rate in proportion to value of nonequilibrium
factor (intensity of stress tensor deviator) with maximum on the
"free" surface on the opening. The second stage passes by means of
microcrack association with production of  randomly oriented
internal surfaces and with rock breakage maximum on the "free"
surface of the opening.

Random character of the internal surfaces
orientation has led to a situation, when among multitude of rock
fragments may be found such, which have acute angle directed in
opposition to the "free" surface. As a consequence with relative
ease they may be displaced in the region of the free space of the
opening under a pressure of another fragments. The availability of
such fragments sharply decreases the stability of rock toward
further fracture. Fracture covers larger and larger region of rock
massive, that sharply reduces net section of the opening and
requires major repairs of it.

The crux of the way next one mustn't wait as long as rock massive
will be collapsed as unwanted, but one may specially destroys it
as necessary. It should be prevented the generation of fragments
described above by means of timely cutting of fragments with such
orientations of its faces, which are favored for further locking.
Moreover, the removal of some part of matter in result of cutting
will have led to its partly or total unloading and to transfer of
maximum of support pressure (and nonequilibrium factor too) in the
region of basements of cut rock fragments. The transfer of maximum
of nonequilibrium factor at certain distance from "free" surface
into rock massive forms conditions for accumulation of microcracks
in this distance and for further macroscopical fracture along
region covering the basements of cut fragments. Such fracture is
favorable to finish separation of cut fragments from the main rock
massive. This is tantamount to creating of a firm rock ring
playing a role of power rock casing for the rest of rock massive.
During fracture in the internal region both loading of the rock
ring will increase and its reverse action on the fracture region
will decrease nonequilibrium factor and as consequence will slow
down the fracture in the whole.

Thus the "partition" of massive by the offered way will have led
to essential (at favor conditions in many times) increasing of
stability and lifetime of a opening. Moreover, at favor conditions
(firm rock, absence of foliation, tectonic dislocations and
cetera) it is possible dispense with traditional synthetic
timbering and to limit with light closed support.
    For execution of way one must cut rock massive around a
cylindrical opening on fragments with some plans. The firsts of
this plans must approximately pass through the axis of the
cylindrical opening, the another are approximately oriented at
right angles to axis (along contour of the opening). The distance
between cuttings, its depth, form and cetera are questions for
principal new technology in the Mining.

\section{Conclusions}

Introduction in the basic ratio (1) thermodynamics of such
internal parameters, as average additional energy per one
microcrack, and density of microcracks, has allowed to give the
physical contents, apparently, to only mechanical task of
destruction of a material and to apply to its analysis standard
methods.

The condition of absence of external sources of
microcracks has allowed to deduce extreme property of free energy
(consisting in equality to zero by its by first by derivative on
internal parameter) from the general positions of thermodynamics,
instead of postulate it as a separate principle.

The equilibrium
values of internal parameter are found and the discontinuity of
the second derivative free energy are determined at transition
through zero of average energy come per microcrack. The equation
of a state (18) for a body with microcracks and expression (19)
for effective elastic modules is received.

\begin{figure}[p]
  \includegraphics[width=7.5in]{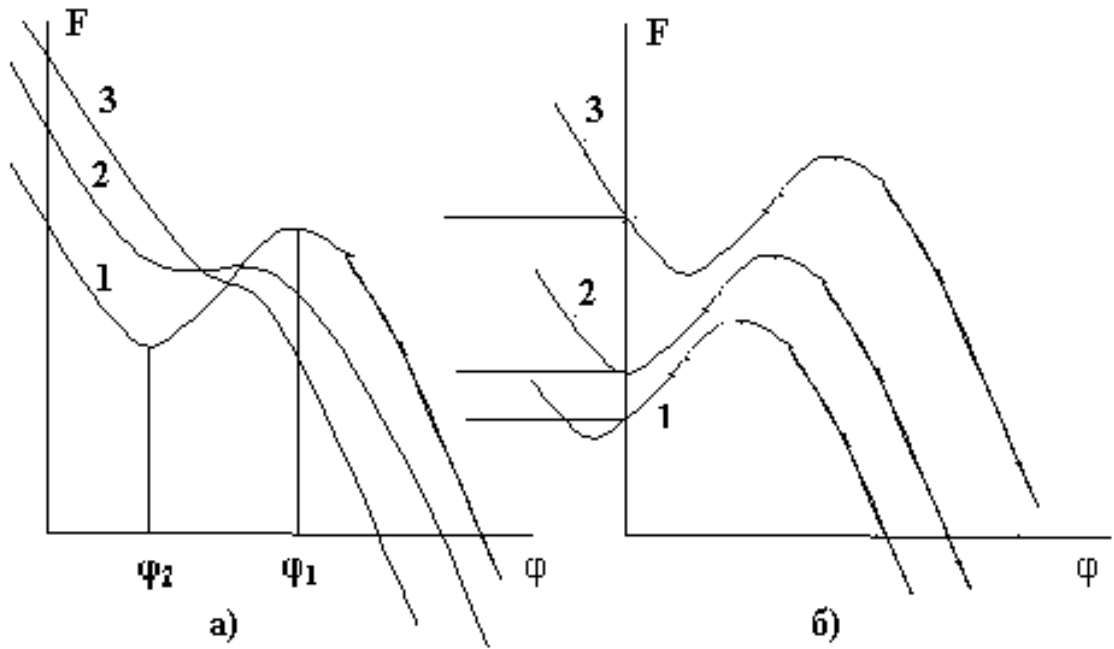}\\
  \caption{Variants of the diagrams of free energy:
a) at various values of discriminant D: 1) - D > 0; 2) - D=0; 3) -
D < 0;  b) at various signs of $\varphi_{2}$: 1) -
$\varphi_{2}<0$; 2) - $\varphi_{2}=0$; 3) - $\varphi_{2}>0$.}\label{f1}
\end{figure}

The developed approach allows to describe process of destruction
of solids from uniform thermodynamic positions. The feature of
destruction of such solids is that under action of mechanical
pressure all kinds genetic discontinuity - microcrack, micropores,
easing etc., always available in any real solid are made active.
At small mechanical influences there is some average size of
microcracks responding an equilibrium state. Since enough large
level of mechanical pressure, the equilibrium state degenerates,
and the system passes in a nonequilibrium mode. In this mode there
is a unlimited growth and merge of microcracks (creep) resulting,
at the end, to macroscopic destruction of a solid. See Figure
\ref{f1}

With the help of a developed formalism it is possible simply
enough to explain the reason essential (under the order of size)
distinction of stability of rock around workins on small and large
depths.
%From the fact of presence of border dividing region of
%various behaviour of the rock the additional estimated connection
%between parameters of the theory is established.

%%% ----------------------------------------------------------------------

% ------------------------------------------------------------------------
%GATHER{Xbib.bib}   % For Gather Purpose Only
%GATHER{Paper.bbl}  % For Gather Purpose Only
\bibliographystyle{amsplain}
\bibliography{probe}
\end{document}